\pdfoutput=1
\documentclass{moriond}

\usepackage{hyperref,amsmath,tikz,multirow,multicol,amsfonts,amssymb,floatrow,blindtext,caption}
\usepackage[compat=1.1.0]{tikz-feynman}
\RequirePackage{luatex85}
\usetikzlibrary{shapes,arrows} 

\newfloatcommand{capbtabbox}{table}[][\FBwidth]

\def\be{\begin{equation}}
\def\ee{\end{equation}}
\def\bea{\begin{eqnarray}}
\def\eea{\end{eqnarray}}

\tikzstyle{decision} = [diamond, draw, fill=yellow!15,
text width=6em, text badly centered, node distance=3cm, inner sep=0pt]
\tikzstyle{block} = [rectangle, draw, fill=blue!15, 
text width=9.5em, text centered, minimum height=2em]
\tikzstyle{block0} = [rectangle, draw, fill=red!15, 
text width=5em, text centered, rounded corners, minimum height=2em]
\tikzstyle{line} = [draw, -latex']
\tikzstyle{cloud1} = [draw, ellipse,fill=green!40, node distance=3cm,
minimum height=2em]
\tikzstyle{cloud2} = [draw, ellipse,fill=red!40, node distance=3cm,
minimum height=2em]
\tikzstyle{io} = [trapezium, trapezium left angle=70, trapezium right angle=110, minimum width=3cm, minimum height=1cm, text centered, draw=black, fill=blue!30]

\newcommand{\Photo}{}

\begin{document}
	\vspace*{4cm}
	\title{Non-Minimal Flavour Violation in $A_4\times SU(5)$ SUSY GUTs}

	\author{\textbf{Samuel J. Rowley}$^1$\footnote{Speaker, contribution to the 2019 EW session of the 54th Rencontres de Moriond}, Jordan Bernigaud$^2$, Bj\"orn Herrmann$^2$, Stephen F. King$^1$}

	\address{
		$^1$School of Physics and Astronomy, Univerisity of Southampton, Southampton, SO17 1BJ, UK\\
		$^2$Univ.\ Grenoble Alpes, USMB, CNRS, LAPTh, 9 Chemin de Bellevue, F-74000 Annecy, France\\
		Email:\texttt{\href{mailto:s.rowley@soton.ac.uk}{s.rowley@soton.ac.uk}}}
	
	\maketitle
	
	\abstracts{
		In these proceedings, we study CP-conserving non-minimal flavour violation in $A_4 \times SU(5)$ inspired Supersymmetric Grand Unified Theories (GUTs), focussing on the regions of parameter space where dark matter is successfully accommodated due to a light right-handed smuon a few GeV heavier than the lightest neutralino. We find that it is necessary to scan over all NMFV parameters simultaneously in order to properly constrain the space of the model.}
	
\section{Introduction}
Despite the absence of experimental evidence, supersymmetric (SUSY) extensions continue to provide attractive solutions to shortcomings of the Standard Model (SM); they cure the hierarchy problem related to the Higgs mass and lead to a more precise gauge-coupling unification as compared to the SM, and can give viable dark matter candidates.

Non-observation of SUSY may to some extent be moderated by the argument that current direct searches rely on specific assumptions. Moreover, as superpartner mass bounds increase, assuming the Minimal Flavour Violation (MFV) paradigm postulating that all flavour-violating interactions are related to the CKM- and PMNS-matrices only, may be relaxed without violating experimental limits. Allowing for additional sources of flavour violation leads to a modification of superpartner decay patterns, hence the obtained mass limits may be weakened \cite{NMFVexp2018}. It appears that a considerable region of the parameter space of the TeV-scale Minimal Supersymmetric Standard Model (MSSM) can accomodate such Non-Minimal Flavour Violation (NMFV) in the squark sector with respect to current experimental and theoretical constraints \cite{Kowalska2014,NMFV2015}.

The goal of this study constrain the NMFV framework of a known model \cite{A4SU5} by introducing off-diagonal squark and slepton mass-squared terms in the Lagrangian at GUT scale, motivated by analyses \cite{Antusch:2013wn,Dimou:2015yng} which show that such flavour violation is generically expected. Here, we take a phenomenological approach, and simply introduce flavour violating terms at high energy to explore their effect on low scale observables. 

\section{Non-Minimal Flavour Violation}
\subsection{Flavour in SUSY-Breaking}
It is well known that Supersymmetry (SUSY) must be broken to some degree. The associated SUSY-breaking Lagrangian contains all terms which do not necessarily respect SUSY but hold to the tenets of gauge invariance and renormalisability. In the MSSM, this reads:
\begin{align}
\begin{split}
\mathcal{L}^{\rm MSSM}_{\rm soft} = 
& - \frac{1}{2} \big( M_1\widetilde{B}\widetilde{B}+M_2\widetilde{W}\widetilde{W} 
+ M_3\widetilde{g}\widetilde{g} + \rm{h.c.} \big) \\[1ex]
& - M_Q^2\widetilde{Q}^{\dagger}\widetilde{Q} - M_L^2\widetilde{L}^{\dagger}\widetilde{L}
- M_U^2\widetilde{U}^*\widetilde{U} - M_D^2\widetilde{D}^*\widetilde{D} 
- M_E^2\widetilde{E}^*\widetilde{E} \\[1ex]
& - \big( A_U \widetilde{U}^*H_u\widetilde{Q} + A_D\widetilde{D}^*H_d\widetilde{Q}
+ A_E \widetilde{E}^*H_d\widetilde{L} + \rm{h.c.} \big) \\[1ex]
& - m_{H_u}^2 H_u^*H_u - m_{H_d}^2 H_d^*H_d - \big( b H_u^*H_d+{\rm h.c.} \big) \,.
\end{split}
\label{Eq:Lagrangian}
\end{align}
While the soft mass and trilinear parameters appearing in Eq.\ \eqref{Eq:Lagrangian} are assumed to be diagonal matrices in flavour space within the MFV framework, they may comprise non-diagonal entries when relaxing this hypothesis and considering a NMFV scenario. It is convenient to parametrize flavour violation in a dimensionless manner by normalising to respective diagonal entries of the sfermion mass matrices;
\begin{gather}
(\delta^Q_{LL})_{ij} = \frac{(M_Q^2)_{ij}}{(M_Q)_{ii}(M_Q)_{jj}},\quad(\delta^U_{RR})_{ij} = \frac{(M_U^2)_{ij}}{(M_U)_{ii}(M_U)_{jj}} \,,\quad
(\delta^D_{RR})_{ij} = \frac{(M_D^2)_{ij}}{(M_D)_{ii}(M_D)_{jj}} \,,\nonumber\\[0.5ex]
(\delta^U_{RL})_{ij}=\frac{v_u}{\sqrt{2}}\frac{(A_U)_{ij}}{(M_Q)_{ii}(M_U)_{jj}} \,,\quad
(\delta^D_{RL})_{ij} = \frac{v_d}{\sqrt{2}}\frac{(A_D)_{ij}}{(M_Q)_{ii}(M_D)_{jj}} \,, \label{Eq:MSSM_small_deltas}\\[0.5ex]
(\delta^L_{LL})_{ij} = \frac{(M_L^2)_{ij}}{(M_L)_{ii}(M_L)_{jj}},\quad (\delta^E_{RR})_{ij} = \frac{(M_E^2)_{ij}}{(M_E)_{ii}(M_E)_{jj}} \,,\quad
(\delta^E_{RL})_{ij} = \frac{v_d}{\sqrt{2}}\frac{(A_E)_{ij}}{(M_L)_{ii}(M_E)_{jj}} \,.\nonumber
\end{gather}
with $v_u$ and $v_d$ being the vacuum expectation values of the up- and down-type Higgs doublets, respectively.
\subsection{The $A_4\times SU(5)$ Model}
We impose $A_4$ and $SU(5)$ symmetries at the GUT scale. To this end, we unify the three families of the usual $F = {\bf \bar{5}} = (d^c, L)$ into the triplet of $A_4$ leading to a unified soft mass parameter $m_F$ for the three generations. Families of $T_i = {\bf 10}_i = (Q, u^c, e^c)_i$ are singlets of $A_4$, and each generation may have an independent soft parameter $m_{T_1}$, $m_{T_2}$, $m_{T_3}$ \cite{Antusch:2013wn}.

Breaking $A_4$ forces off-diagonal elements to be smaller than diagonal entries, providing a theoretical motivation for small-but-non-zero flavour violation in such a class of models. $SU(5)$ gives the following relationships between the dimensionless NMFV parameters in the basis before rotation to the SCKM basis, at the GUT scale:
\begin{align}
	\begin{split}
		\delta^{Q_0}_{LL} ~=~ \delta^{U_0}_{RR} ~=~  \delta^{E_0}_{RR} ~&\equiv~ \delta_{TT}, \\	
		\quad\delta^{D_0}_{RR} ~=~ \delta^{L_0}_{LL} ~&\equiv~ \delta_F \,, \\
		\delta^{D_0}_{RL} ~=~ (\delta^{E_0}_{RL})^T ~&\equiv~ \delta_{FT} \,,\\
		\delta^{U_0}_{RL} ~&\equiv~ \delta_{TT} \,.
		\label{Eq:NMFV_GUT_A4xSU5_deltas}
	\end{split}
\end{align}
These four matrices parametrize the flavour violation in the $A_4\times SU(5)$ setup studied here. Note that $\delta_T$, $\delta_F$ and $\delta_{TT}$ are necessarily symmetric whereas $\delta_{FT}$ is not leading to a total of 15 NMFV parameters at the GUT scale.

It is apparent that we have flavour violation at phenomenological scales from two sources; the presence of off-diagonal elements in various coupling matrices at the GUT scale due to $A_4$ breaking, and further effects induced by RGE running.

\section{Setup and Tools}
We consider two MFV reference parameter points, one of which is inspired by a previous study of this model \cite{A4SU5}, and the other one with a heavier smuon. In both cases, we switch on off-diagonal mass terms, consistent with $SU(5)$, arising from $A_4$ breaking effects.

Diagonal entries of soft matrices are fixed and NMFV parameters are entered for each point by random selection about empirically determined limits. Parameters are then handed to \texttt{SPheno}\cite{SARAH2014} at the GUT scale, and are run using two-loop RGEs down to low scales where data is available for comparison. Flavour phenomena as listed in table \ref{tab:exp_data} are calculated using \texttt{SPheno} and the relic density of the lightest neutralino (our DM candidate) is calculated by a custom version of \texttt{micrOMEGAs}\cite{micrOMEGAs2016}.

Predictions are then compared against experimental data to determine if the point under test is viable \cite{PDG2018,HFLAF2017}. In this manner of comparison, we set upper bounds on the amount of flavour violation allowed in this scenario. It is necessary to scan over all flavour violating parameters simultaneously, as to exploit hidden correlations and cancellations, and obtain accurate limits on flavour violation.
\begin{figure}
	\begin{floatrow}
		\centering
		\ffigbox{
		\begin{tikzpicture}[node distance = 1.2cm]
		\begin{scope}[scale=0.58, transform shape]
		\node [cloud1] (init) {MFV and NMFV Params};
		\node [block,below of=init] (mats) {SPhenoMSSM-4.0.3};
		\node [decision,below of=mats] (phys) {Physical Spectrum, Neutral LSP?};
		\node [block0,right of=phys, xshift=2.5cm] (exclude1) {Point Excluded};
		\node [block,below of=phys,yshift=-2.0cm] (micromegas) {micrOMEGAs-4.3.5};
		\node [decision,below of=micromegas] (constcheck) {Constraint Checks};
		\node [block0,right of=constcheck,xshift=2.5cm] (exclude2) {Prior Only};
		\node [cloud1, below of=constcheck] (predictions) {Prior and Posterior Distributions};
		\node [io, above of=init] (rand scan) {Overarching Flat Random Scan};
		
		\node [left of=rand scan] (left1) {};
		\node [left of=init] (left2) {} ; 
		
		\path [line] (init) -- (mats);
		\path [line] (mats) -- (phys);
		\path [line] (micromegas) -- (constcheck);
		\draw [dashed, blue] (rand scan) -- (init);
		
		\draw [line] (phys) -- node [anchor=south] {no} (exclude1);
		\draw [line] (phys) -- node [anchor=east] {yes} (micromegas);
		\draw [line] (constcheck) -- node [anchor=south] {fail} (exclude2);
		\draw [line] (constcheck) -- node [anchor=east] {pass} (predictions);
		\end{scope}
		\end{tikzpicture}
		}
		{\caption{Proceedure for each parameter point}
		\label{fig:flowchart}}
		\capbtabbox{
			\renewcommand{\arraystretch}{1.25}
			\scalebox{0.85}{
				\begin{tabular}{|c|c|}
					\hline
					Observable & Constraint\\
					\hline
					$m_h$ & $\left( 125.2 \pm 2.5 \right)$ GeV\\
					\hline
					$\mathrm{BR}(\mu \rightarrow e\gamma)$ & $ <4.2 \times 10^{-13}$\\
					$\mathrm{BR}(\mu \rightarrow 3e)$ & $ <1.0 \times 10^{-12}$\\
					$\mathrm{BR}(\tau \rightarrow e\gamma)$ & $ <3.3 \times 10^{-8}$\\
					$\mathrm{BR}(\tau \rightarrow \mu\gamma)$ & $ <4.4 \times 10^{-8}$\\
					$\mathrm{BR}(\tau \rightarrow 3e)$ & $ <2.7 \times 10^{-8}$\\
					$\mathrm{BR}(\tau \rightarrow 3\mu)$ & $ <2.1 \times 10^{-8}$\\
					$\mathrm{BR}(\tau \rightarrow e^-\mu\mu)$ & $ <2.7 \times 10^{-8}$\\
					$\mathrm{BR}(\tau \rightarrow e^+ \mu \mu)$ & $ <1.7 \times 10^{-8}$ \\
					$\mathrm{BR}(\tau \rightarrow \mu^- e e)$ & $ <1.8 \times 10^{-8}$\\
					$\mathrm{BR}(\tau \rightarrow \mu^+ e e)$ & $ <1.5 \times 10^{-8}$\\
					\hline
					$\mathrm{BR}(B \rightarrow X_s \gamma)$ & $\left( 3.32 \pm 0.18 \right) \times 10^{-4}$\\
					$\mathrm{BR}(B_s \rightarrow \mu \mu)$ & $\left( 2.7 \pm 1.2 \right) \times 10^{-9}$\\
					$\Delta M_{B_s}$ & $\left( 17.757 \pm 0.312 \right)$ ps$^{-1}$\\
					$\Delta M_K$ & $\left( 3.1 \pm 1.2 \right)\times10^{-15} $ GeV\\
					$\epsilon_K$ & $2.228 \pm 0.29$\\
					\hline
					$\Omega_{\rm DM}h^2$ & $0.1198 \pm 0.0042$\\
					\hline
				\end{tabular}
			}}
		{\caption{Data used to constrain parameters}
		\label{tab:exp_data}}
	\end{floatrow}
\end{figure}

\section{Results}
\begin{figure}
	\centering
	\includegraphics[width=0.46\linewidth]{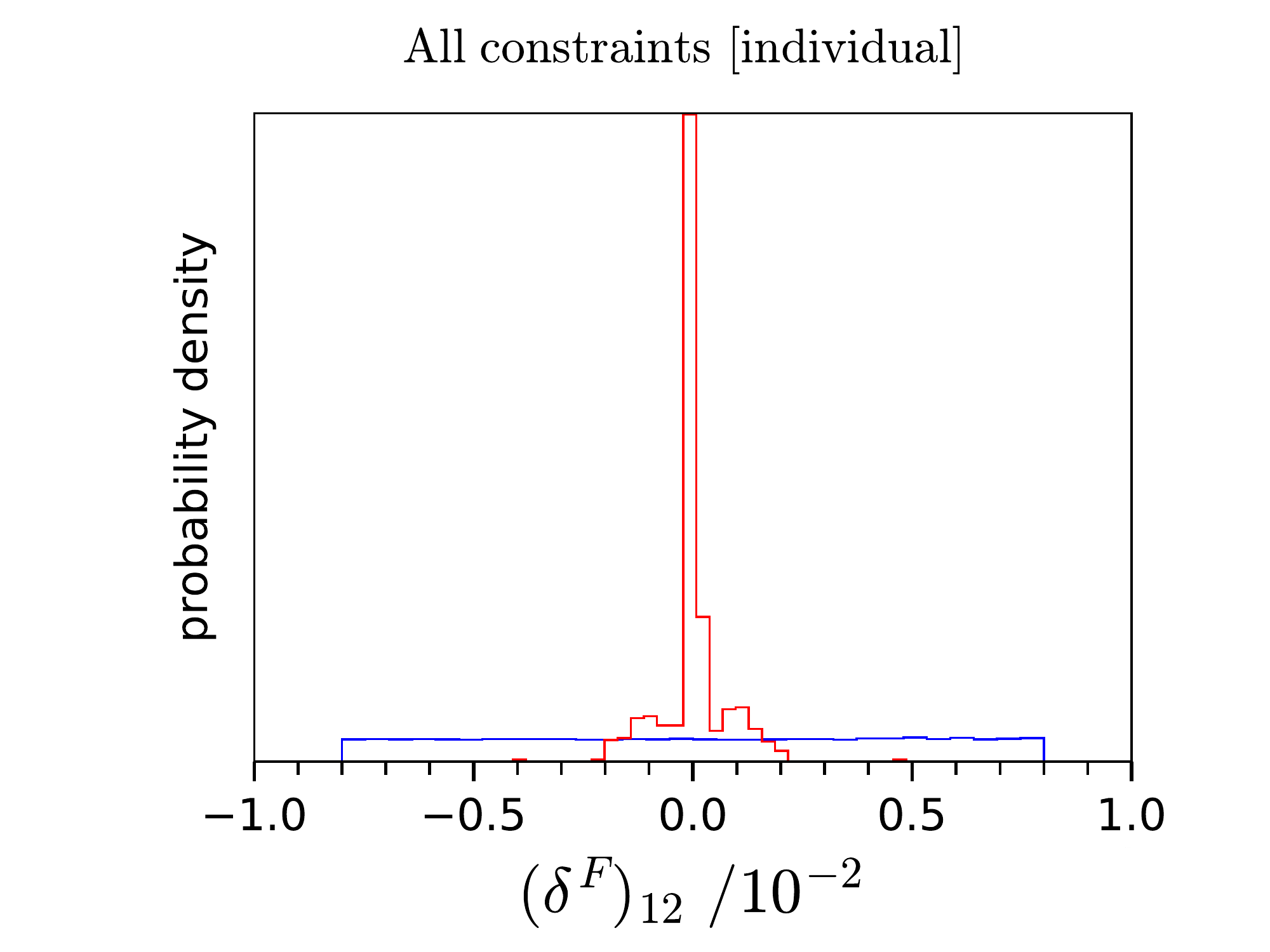}
	\includegraphics[width=0.46\linewidth]{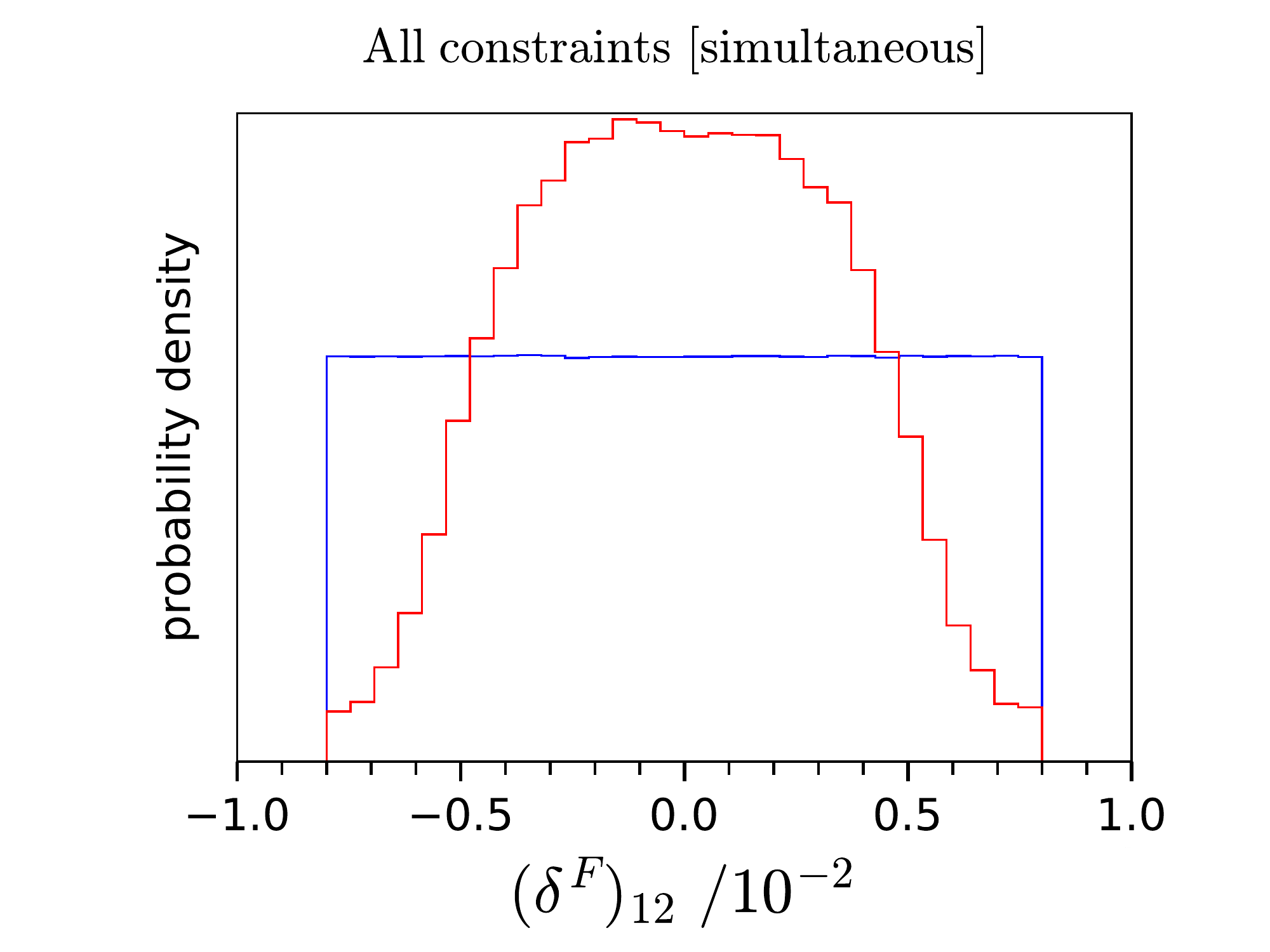}
	\caption{Constraints on a particular flavour violating parameter. Prior distributions shown in blue, posteriors in red. Left panel shows constraint when parameter scanned over in isolation, right panel when scanned over with all other parameters.}
	\label{fig:costraint_plot}
\end{figure}

Using the proceedure outlined above, we constrain the NMFV parameter space of this model in both reference scenarios. In figure \ref{fig:costraint_plot}, we demonstrate the importance of scanning over multiple parameters in tandem; the left panel details the result of allowing a single parameter to vary, keeping all others fixed. In such a case the posterior distribution (in red) is \emph{sharply} peaked around 0, meaning that not much flavour violation is allowed.

In a stark contrast to this, the right panel of figure \ref{fig:costraint_plot} shows a broad distribution, allowing a considerable amount of flavour violation.
For a detailed discussion of all results, we encourage the reader to consult the full paper \cite{Bernigaud2018}.
\section{Conclusion}
 As experiments continue to exlcude parameter space for the MSSM and other favoured minimal SUSY theories, it is important to investigate and constrain those more exotic scenarios that may include non-universal gaugino masses, compressed spectra etc. that may be able to elude conventional collider searches.
 
 We have constrained the flavour violating parameter space of the MSSM in this specific GUT scenario. We stress that small flavour violation is a prediction of unified models that include a flavour symmetry, and as such studying how superpartners can influence various flavour violating phenomena is a critical part of testing SUSY at currently accessible scales.
%
 
 Work in this area is ongoing, with an eye to enable SUSY GUT model descrimination using flavour physics and to further constrain the MSSM.


\section{References}
\begin{thebibliography}{99}

\bibitem{NMFVexp2018}
  A.~Chakraborty, M.~Endo, B.~Fuks, B.~Herrmann, M.~M.~Nojiri, P.~Pani and G.~Polesello,
  Eur.\ Phys.\ J.\ C {\bf 78} (2018) no.10,  844
  doi:10.1140/epjc/s10052-018-6331-x
  [arXiv:1808.07488 [hep-ph]].

\bibitem{Kowalska2014}
  K.~Kowalska,
  JHEP {\bf 1409} (2014) 139
  doi:10.1007/JHEP09(2014)139
  [arXiv:1406.0710 [hep-ph]].
          
\bibitem{NMFV2015}
  K.~De Causmaecker, B.~Fuks, B.~Herrmann, F.~Mahmoudi, B.~O'Leary, W.~Porod, S.~Sekmen and N.~Strobbe,
  JHEP {\bf 1511} (2015) 125
  doi:10.1007/JHEP11(2015)125
  [arXiv:1509.05414 [hep-ph]].
 
\bibitem{A4SU5}
  A.~S.~Belyaev, S.~F.~King and P.~B.~Schaefers,
  Phys.\ Rev.\ D {\bf 97} (2018) no.11,  115002
  doi:10.1103/PhysRevD.97.115002
  [arXiv:1801.00514 [hep-ph]].

\bibitem{Antusch:2013wn}
	S.~Antusch, S.~F.~King and M.~Spinrath,
	Phys.\ Rev.\ D {\bf 87} (2013) no.9,  096018
	doi:10.1103/PhysRevD.87.096018
	[arXiv:1301.6764 [hep-ph]].

\bibitem{Dimou:2015yng}
  M.~Dimou, S.~F.~King and C.~Luhn,
  JHEP {\bf 1602} (2016) 118
  doi:10.1007/JHEP02(2016)118
  [arXiv:1511.07886 [hep-ph]].
  
\bibitem{SARAH2014}
W.~Porod, F.~Staub and A.~Vicente,
Eur.\ Phys.\ J.\ C {\bf 74} (2014) no.8,  2992
doi:10.1140/epjc/s10052-014-2992-2
[arXiv:1405.1434 [hep-ph]].

\bibitem{micrOMEGAs2016}
D.~Barducci, G.~B\'elanger, J.~Bernon, F.~Boudjema, J.~Da Silva, S.~Kraml, U.~Laa and A.~Pukhov,
Comput.\ Phys.\ Commun.\  {\bf 222} (2018) 327
doi:10.1016/j.cpc.2017.08.028
[arXiv:1606.03834 [hep-ph]].

\bibitem{PDG2018}
 M.~Tanabashi {\it et al.} [Particle Data Group],
 Phys.\ Rev.\ D {\bf 98} (2018) 030001.

\bibitem{HFLAF2017} 
  Y.~Amhis {\it et al.} [HFLAV Collaboration],
  Eur.\ Phys.\ J.\ C {\bf 77}, no. 12, 895 (2017)
  doi:10.1140/epjc/s10052-017-5058-4
  [arXiv:1612.07233 [hep-ex]], and online update at  \href{https://hflav.web.cern.ch}{{\texttt{https://hflav.web.cern.ch}}}.

\bibitem{Bernigaud2018}
	J.~Bernigaud, B.~Herrmann, S.~F.~King and S.~J.~Rowley.
	JHEP \textbf{1903} (2019) 067
	doi:10.1007/JHEP03(2019)067
	[arXi:1812.07463 [hep-ph]].
\end{thebibliography}
\end{document}